\documentclass[aps,pra,reprint,superscriptaddress,longbibliography]{revtex4-1}
\usepackage[unicode=true,pdfusetitle, bookmarks=true,bookmarksnumbered=false,bookmarksopen=false, breaklinks=false,pdfborder={0 0 0},backref=false,colorlinks=false] {hyperref}
\hypersetup{colorlinks,linkcolor=myurlcolor,citecolor=myurlcolor,urlcolor=myurlcolor}
\usepackage{braket,colortbl,amsthm,amsmath,amssymb,txfonts,graphicx}
\definecolor{myurlcolor}{rgb}{0,0,0.7}
\usepackage{float}
\usepackage{cleveref}
\usepackage{xcolor}
\usepackage{subfigure}

\theoremstyle{plain}
\usepackage{mathrsfs}
\usepackage{comment}
\usepackage{ragged2e}  

\usepackage[utf8]{inputenc}
\usepackage[normalem]{ulem}
\usepackage{float}
\usepackage{graphicx,subcaption,ragged2e}
\usepackage{mathrsfs}
\usepackage{braket}
\usepackage{physics}
\usepackage{mwe}
\usepackage{ragged2e}\usepackage{pifont}

\DeclareMathAlphabet{\mathbcal}{OMS}{cmsy}{b}{n}

\usepackage{tikz}
\usepackage[margin=1in]{geometry}
\usepackage{ragged2e}
\usetikzlibrary{decorations.pathmorphing,arrows.meta,shapes.symbols,3d,calc}

\theoremstyle{remark}

\begin{document}

\title{Relativistic Quantum Thermal Machine: Harnessing Relativistic Effects to Surpass Carnot Efficiency}

\author{Tanmoy Pandit}
 \email{tanmoypandit163@gmail.com}
 \affiliation{Institute for Theoretical Physics, Leibniz Institute of Hannover, Hannover, Germany}
\affiliation{Institute of Physics and Astronomy, TU Berlin, Berlin, Germany}

\author{Pritam Chattopadhyay}
\email{pritam.chattopadhyay@weizmann.ac.il}
\affiliation{Department of Chemical and Biological Physics,
Weizmann Institute of Science, Rehovot 7610001, Israel}

\author{Kaustav Chatterjee}
\email{kauch@dtu.dk}
\affiliation{Center for Macroscopic Quantum States (bigQ), Department of Physics, Technical University of Denmark, \\
 Building 307, Fysikvej, 2800 Kongens Lyngby, Denmark}

\author{Varinder Singh}
\email{varinderkias@kias.re.kr}
\affiliation{School of Physics, Korea Institute for Advanced Study, Seoul 02455, Korea}

\begin{abstract}
We investigate a three-level maser quantum thermal machine in which the system–reservoir interaction is modeled via Unruh–DeWitt type coupling, with one or both reservoirs undergoing relativistic motion relative to the working medium. Motion induces Doppler reshaping of the reservoir spectra, modifying energy-exchange rates and enabling operation beyond the Carnot efficiency at finite power. We numerically analyze families of efficiency–power curves and extract the analytic form of a generalized Carnot bound, which recovers the Carnot limit. In addition, Doppler reshaping alters the boundaries between heat-engine and refrigerator operation, making it possible to extract positive work even in the absence of a temperature gradient. These findings establish relativistic motion as a genuine thermodynamic resource.

\end{abstract}
\maketitle

\textit{Introduction--} Since the inception of quantum thermodynamics, quantum heat engines (QHEs) have served as prototypical models for extending classical thermodynamic principles to quantum the quantum regime ~\cite{PhysRevLett.2.262,scully_science,PhysRevE.49.3903,PhysRevLett.98.240601,PhysRevE.77.041118,chattopadhyay2021quantum,PhysRevLett.109.203006,chattopadhyay2020non,pandit2021non,santos2023pt,sur2024many,Bera12024,mohan2025coherent,pandit2022bounds,green2022probing,reviewPr,mohan2025enhancing}. Initiated by the pioneering realization that a \textit{three-level maser} could operate as a continuous heat engine~\cite{PhysRevLett.2.262}, the field has grown to explore the roles of coherence and correlation~\cite{scully_pnas,VSTUR2023}, many-body effects~\cite{PhysRevLett.124.210603}, quantum uncertainty~\cite{chattopadhyay2021bound}, and finite-time cycles~\cite{PhysRevLett.119.050601}, 
%
with implementations ranging from trapped ions~\cite{PhysRevLett.109.203006}, atomic clouds and transmons~\cite{cherubim2019non}, quantum dots~\cite{van2020single}, with several experimental realizations already demonstrated~\cite{bouton2021quantum}. 
\\\\
Meanwhile, inspired by recent advances in nanotechnology~\cite{RevModPhys.86.1391} and the emergence of Dirac materials like graphene and Weyl semimetals~\cite{RevModPhys.81.109}, there is growing interest in relativistic QHEs, with studies proposing thermodynamic cycles for relativistic particles~\cite{PhysRevE.94.022109,PhysRevE.86.061108,chattopadhyay2019relativistic, ferketic2023boosting} and heat engines based on Dirac systems~\cite{PhysRevE.91.052152}, bringing once-speculative ideas within experimental reach. While relativistic engines, such as relativistic Otto cycles, have been studied extensively \cite{myers2021quantum,HirotaniGallockYoshimura2025,gallock2024relativistic,kane2021entangled}, the thermodynamic behavior of continuous, autonomous relativistic heat engines remains largely unexplored. At the same time, the question of whether \textit{relativistic motion can be harnessed as a genuine thermodynamic resource} allowing performance beyond conventional thermal settings remains unresolved.
\\\\ 
In this work, we address this critical gap by investigating a continuous three-level maser QHE where one or both thermal reservoirs undergo relativistic motion with constant velocity with respect to the working medium. Our approach employs an \textit{Unruh–DeWitt} (UDW) type coupling, which has been widely used to model light–matter interactions in relativistic quantum field theory. This framework naturally captures how the working medium exchanges energy with moving thermal reservoirs and allows us to treat relativistic motion on the same footing as thermal fluctuations  \cite{Unruh1976,DeWitt1979,Costa1995}. The choice of the model is motivated by the fact that quantum heat engines have already been realized in this setting \cite{Klatzow2019,Zou2017}, making it an experimentally viable and fully autonomous platform for exploring relativistic effects.\\\\
Our central findings are twofold. First, we show that the modified energy exchange due to relativistic effects enables the engine to operate beyond the Carnot efficiency, and we also found an analytic expression for this generalized Carnot bound. Second, we demonstrate that the work output depends non-monotonically on reservoir motion, with pronounced asymmetries between moving-hot and moving-cold configurations. This behavior shifts the boundaries between engine and refrigerator operation and even enables positive work extraction when the two reservoirs are at the same temperature, thereby establishing relativistic motion as a genuine thermodynamic resource.

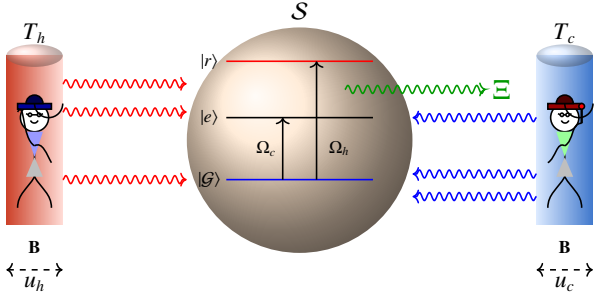
\begin{figure}[H]
  \centering
  \resizebox{\linewidth}{!}{
    \begin{tikzpicture}[thick, every node/.style={scale=1.0}]

    \definecolor{hotred}{RGB}{255,80,50}
    \definecolor{coldblue}{RGB}{80,150,255}

    \shade[left color=hotred!80!black, right color=hotred!20!white] 
        (-5.2,1.5) rectangle (-4.2,-1.5);
    \shade[ball color=hotred!40!white, opacity=0.5] 
        (-5.2,1.5) arc[start angle=180, end angle=360, x radius=0.5, y radius=0.2] --
        (-4.2,1.5) arc[start angle=0, end angle=180, x radius=0.5, y radius=0.2];

    \draw[fill=white] (-4.7,0.4) circle(0.25); 
    \draw[fill=black!40!blue] (-5.0,0.55) rectangle (-4.4,0.65); 
    \draw[fill=black!70!blue] (-4.85,0.65) arc[start angle=180, end angle=0, radius=0.15] -- (-4.7,0.65) -- cycle; 

    \draw[thick] (-4.78,0.45) circle(0.03); 
    \draw[thick] (-4.62,0.45) circle(0.03); 
    \draw[thick] (-4.75,0.45) -- (-4.65,0.45); 
    \draw[thick] (-4.81,0.45) -- (-4.85,0.47); 
    \draw[thick] (-4.59,0.45) -- (-4.55,0.47); 
    \draw[thick] (-4.77,0.35) arc[start angle=200, end angle=340, radius=0.07]; 
    \draw[thick] (-4.7,0.65) arc[start angle=180, end angle=360, radius=0.25]; 

    \draw[thick] (-4.7,0.15) -- (-4.7,-0.65); 

    \fill[blue!40] (-4.85,0.15) -- (-4.55,0.15) -- (-4.7,-0.3) -- cycle;

    \fill[gray!50] (-4.7,-0.3) -- (-4.85,-0.65) -- (-4.55,-0.65) -- cycle;

    \draw[thick] (-4.7,-0.1) to[out=150,in=60] (-5.0,0.5); 
    \draw[thick] (-4.7,-0.1) to[out=30,in=120] (-4.4,-0.4); 

    \draw[thick] (-4.7,-0.65) to[out=-120,in=90] (-5.0,-1.2); 
    \draw[thick] (-4.7,-0.65) to[out=-60,in=90] (-4.4,-1.2);  

    \node at (-4.7,1.9) {\large $T_h$};
    \node at (-4.7,-1.9) {\textbf{B}};
    \draw[<->, dashed] (-5.2,-2.3) -- (-4.2,-2.3) node[midway, below] {\large $u_h$};

    \shade[right color=coldblue!80!black, left color=coldblue!20!white] 
        (4.2,1.5) rectangle (5.2,-1.5);
    \shade[ball color=coldblue!40!white, opacity=0.5] 
        (4.2,1.5) arc[start angle=180, end angle=360, x radius=0.5, y radius=0.2] --
        (5.2,1.5) arc[start angle=0, end angle=180, x radius=0.5, y radius=0.2];

    \draw[fill=white] (4.7,0.4) circle(0.25); 
    \draw[fill=black!40!red] (4.4,0.55) rectangle (5.0,0.65); 
    \draw[fill=black!70!red] (4.55,0.65) arc[start angle=180, end angle=0, radius=0.15] -- (4.7,0.65) -- cycle; 

    \draw[thick] (4.65,0.45) circle(0.03); 
    \draw[thick] (4.75,0.45) circle(0.03); 
    \draw[thick] (4.68,0.45) -- (4.72,0.45); 
    \draw[thick] (4.62,0.45) -- (4.58,0.47); 
    \draw[thick] (4.78,0.45) -- (4.82,0.47); 
    \draw[thick] (4.64,0.35) arc[start angle=200, end angle=340, radius=0.07]; 
    \draw[thick] (4.7,0.65) arc[start angle=180, end angle=360, radius=0.25]; 

    \draw[thick] (4.7,0.15) -- (4.7,-0.65); 

    \fill[green!40] (4.55,0.15) -- (4.85,0.15) -- (4.7,-0.3) -- cycle;

    \fill[gray!50] (4.7,-0.3) -- (4.55,-0.65) -- (4.85,-0.65) -- cycle;

    \draw[thick] (4.7,-0.1) to[out=150,in=30] (4.4,-0.4);   
    \draw[thick] (4.7,-0.1) to[out=30,in=180] (5.0,0.3);    
    \draw[very thick] (5.0,0.3) -- (5.0,0.6); 
    \draw[fill=red] (5.0,0.6) circle(0.05);

    \draw[thick] (4.7,-0.65) to[out=-120,in=90] (4.4,-1.2);  
    \draw[thick] (4.7,-0.65) to[out=-60,in=90] (5.0,-1.2);   

    \node at (4.7,1.9) {\large $T_c$};
    \node at (4.7,-1.9) {\textbf{B}};
    \draw[<->, dashed] (4.2,-2.3) -- (5.2,-2.3) node[midway, below] {\large $u_c$};

    \shade[ball color=orange!20] (0,0) circle (2cm);
    \node at (0,2.3) {\large $\mathcal{S}$};

    \draw[blue,  thick] (-1.3,-0.7) -- (1.3,-0.7);
    \node at (-1.6,-0.7) {$|\mathcal{G}\rangle$};
    \draw[black, thick] (-1.3,0.4) -- (1.3,0.4);
    \node at (-1.6,0.4) {$|e\rangle$};
    \draw[red,   thick] (-1.3,1.4) -- (1.3,1.4);
    \node at (-1.6,1.4) {$|r\rangle$};

    \draw[->, thick] (-0.3,-0.7) -- (-0.3,0.4) node[midway, left] {$\Omega_c$};
    \draw[->, thick] (0.3,-0.7) -- (0.3,1.4) node[pos=0.35, right=3pt, yshift=-5pt] {$\Omega_h$};

    \draw[decorate, decoration={snake, amplitude=2pt, segment length=5pt}, red, thick, ->]
        (-4.2,1.0) -- (-2.0,1.0);
    \draw[decorate, decoration={snake, amplitude=2pt, segment length=5pt}, red, thick, ->]
        (-4.2,0.5) -- (-2.0,0.5);
    \draw[decorate, decoration={snake, amplitude=2pt, segment length=5pt}, red, thick, ->]
        (-4.2,-0.7) -- (-2.0,-0.7);

    \draw[decorate, decoration={snake, amplitude=2pt, segment length=5pt}, blue, thick, ->]
        (4.2,-0.6) -- (2.0,-0.6);
    \draw[decorate, decoration={snake, amplitude=2pt, segment length=5pt}, blue, thick, ->]
        (4.2,-1.0) -- (2.0,-1.0);
    \draw[decorate, decoration={snake, amplitude=2pt, segment length=5pt}, blue, thick, ->]
        (4.2,0.4) -- (2.0,0.4);

    \draw[decorate, decoration={snake, amplitude=2pt, segment length=5pt}, green!60!black, thick, ->]
        (0.8,0.9) -- ++(2.5,0) node[right] {\Large $\Xi$};

    \end{tikzpicture}
  }
  \caption{\justifying Schematic of a three-level thermal machine coupled to two thermal reservoirs. $u_h$, $u_c$ denote the rapidity of the non-stationary reservoirs with temperatures $T_h$ and $T_c$ respectively. $|\mathcal{G}\rangle$ is the ground state, the excited state is $|r \rangle$, and $|e\rangle$ denotes the intermediate state. $\Xi$ represents the strength of matter-field coupling.}
  \label{fig:RQHE}
\end{figure}

\color{black}
\textit{Model--}
We consider a three-level maser continuously interacting with two moving thermal reservoirs and a single-mode classical field, as illustrated in Fig.~\ref{fig:RQHE}. The hot bath at temperature $T_h$ couples the ground $|\mathcal{G}\rangle$ and the excited state $|r\rangle$, while the cold bath at temperature $T_c$ couples the ground $|\mathcal{G}\rangle$ and the intermediate state $|e\rangle$. The transition $|e\rangle \leftrightarrow |r\rangle$ is coherently driven by a classical field of frequency $\Omega$, providing the energy bias necessary for continuous power extraction. 
The bare Hamiltonian of the three-level system is  
$
   \hat{\mathcal{H}}_0=\sum_{j=\{\mathcal{G},e,r\}}\Omega_j\,\ket{j}\bra{j}$, 
   where $\Omega_j$ denotes the energy of the level $\ket{j}$. The interaction with the classical driving field is described, within the rotating-wave approximation~\cite{BoukobzaTannor2007}, by
$
   \hat{\mathcal{V}}(t)=\Xi\big(e^{-i\Omega t}\ket{r}\bra{e}+\mathrm{H.c.}\big),
$
where $\Xi$ is the matter--field coupling constant and $\Omega=\Omega_r-\Omega_{\mathcal{G}}$.

To describe these system–reservoir interactions in the relativistic regime, we model each thermal bath as a quantum field in a thermal state and employ UDW type coupling ~\cite{Unruh1976,DeWitt1979}. This framework offers a relativistically consistent way to capture how the moving reservoirs exchange energy with the working medium: relativistic motion reshapes the spectral distribution of the baths through Doppler effects, thereby modifying transition rates and ultimately influencing the engine’s performance.

In this work, we focus on reservoirs undergoing constant-velocity ($v$) trajectories parameterized by the rapidity $u$. Specifically, we consider worldlines of the form \cite{PapadatosA,PapadatosB}
\begin{equation}
    x(\tau) = (\cosh{u},\sinh{u},0,0)\tau, 
\end{equation}
where $\tau$ is the proper time and the associated velocity is $v=\tanh u$. This parametrization provides a direct and compact link between relativistic motion and the observed bath spectrum in the system’s rest frame. For such trajectories, the reduced dynamics of the matter–field system is described by the  Lindblad master equation in the rotating frame with respect to $\hat{\mathcal{H}}_0$ \cite{PapadatosA,PapadatosB}:
\begin{equation}\label{eq:master}
    \dot{\rho}=-i\,[ \hat{\mathcal{V}_R},\rho]
    +\mathcal{L}_h^{\mathrm{rel}}[\rho]
    +\mathcal{L}_c^{\mathrm{rel}}[\rho],
\end{equation}
where the dissipators take the form
\begin{align}
\mathcal{L}_h^{\mathrm{rel}}[\rho]&=
\Gamma_h\big(N_h(u_h)+1\big)\,\mathcal{D}[\sigma_{\mathcal{G}r}]\,\rho
+\Gamma_h N_h(u_h)\,\mathcal{D}[\sigma_{r\mathcal{G}}]\,\rho, \\
\mathcal{L}_c^{\mathrm{rel}}[\rho]&=
\Gamma_c\big(N_c(u_c)+1\big)\,\mathcal{D}[\sigma_{\mathcal{G}e}]\,\rho
+\Gamma_c N_c(u_c)\,\mathcal{D}[\sigma_{e\mathcal{G}}]\,\rho,
\end{align}
with $\mathcal{D}[A]\rho = A\rho A^\dagger - \tfrac{1}{2}{A^\dagger A,\rho}$ the standard Lindblad superoperator, $\Gamma_{h,c}$ the \textit{Weisskopf–Wigner} decay constants, and $\sigma_{ab}\equiv\ket{a}\bra{b}$.
The modified photon occupation numbers $N_{h,c}$ are given by
\begin{equation}
N_{u_{h(c)}}(\Omega_{h(c)}) = \frac{1}{2\beta\Omega_{h(c)}\sinh u_{h(c)}}
\ln\left[\frac{1-e^{-\beta\Omega_{h(c)} e^{|u_{h(c)}|}}}{1-e^{-\beta\Omega_{h(c)} e^{-|u_{h(c)}|}}}\right].
\end{equation}
which encode the effect of relativistic motion explicitly through the bath rapidities $u_h$ and $u_c$ \cite{PapadatosA}.

\textit{Quantities of interest--}
Assuming weak coupling between the system and the reservoirs, the heat flux, power output, and efficiency of the SSD engine are defined within the standard framework (working in natural units with $\hbar=1$), following ~\cite{BoukobzaTannor2007}:
\begin{eqnarray}
\dot{Q_h} &=&  {\rm Tr}(\mathcal{L}_h[\rho_R]\hat{\mathcal{H}}_0), \label{heat1} \\
\mathcal{P} &=& i {\rm Tr}([\hat{\mathcal{H_0}},\hat{\mathcal{V}}_R]\rho_R), \label{power1} \\
\eta &=& \frac{P}{\dot{Q_h}}. \label{efficiencycrude}
\end{eqnarray}
Here, we employed the sign convention in which heat absorbed from the hot bath, heat released to the cold bath, and the power output are all taken as positive.   Eq.~\eqref{eq:master} can be solved to yield the steady state solution $\rho_R$ (see Appendix \ref{App. B} for details). Then, by substituting the expressions for $\hat{\mathcal{H}}_0$, $\hat{\mathcal{V}}_R$, and $\mathcal{L}_h[\rho_R]$ into Eqs.~(\ref{heat1})--(\ref{efficiencycrude}), we can obtain the expressions for the power and efficiency of the heat engine. 

The expression for the power can be written in the compact form,
\begin{equation}
\mathcal{P} =   \frac{4(N_h(u_h)-N_c(u_c))\Gamma_h\Gamma_c\Xi^2}{4\Xi^2 B+C\,D\, \Gamma_c\Gamma_h}, \label{power}
\end{equation}
where $B=\Gamma_h(3N_h+1)+\Gamma_c(3N_c+1)$,   $C=3N_h N_c +2N_h+2N_c+1$, and $D=\Gamma_h (N_h+1)+\Gamma_c (N_c+1)$.
Similarly, efficiency of the engine is given by
\begin{equation}
\eta=1-\frac{\Omega_c}{\Omega_h}. \label{eff}
\end{equation}
which retains the same form as in maser heat engines coupled to two stationary reservoirs~\cite{Varinder2020,Dorfman2018}. Thus, relativistic motion modifies the populations and rates but does not alter the structural form of the efficiency.

\begin{figure}
    \centering
    \includegraphics[width=\linewidth]{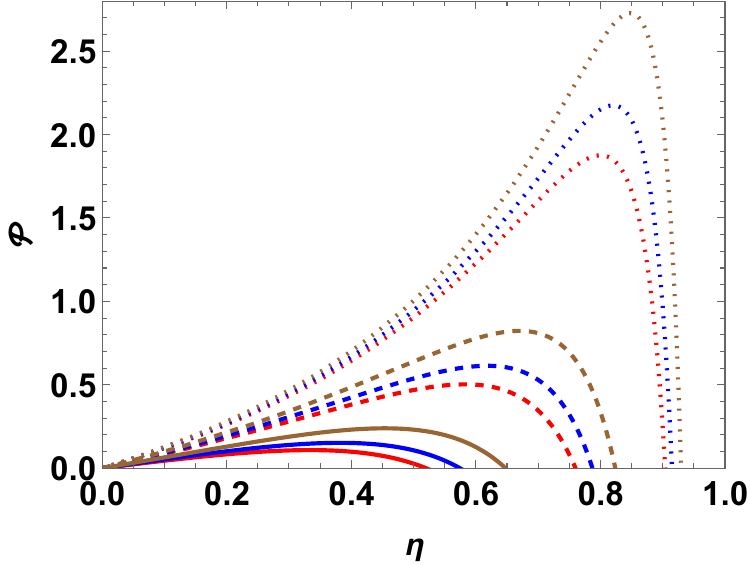}
    \caption{\justifying
    Efficiency–power $(\eta,\mathcal{P})$ curves in the high-temperature limit for representative fixed parameter sets ($u, \tau$). Curves are grouped by temperature ratio $\tau \in \{0.5, 0.25,  0.1\}$     (red, blue, brown, respectively) and, within each group, by relativistic parameter $u \in \{0.5, 1, 1.5\}$  with line styles (solid, dashed, dotted). 
   Each curve exhibits the characteristic $\eta$-$\mathcal{P}$ trade-off; endpoints at $\mathcal{P} \rightarrow 0$ mark the maximum efficiency attainable for that set. The envelope of endpoints yields the generalized bound $\eta_{\rm up} = 1 - \tau \, u / \sinh u$ [Eq. (\ref{upperbound})]. }
    \label{fig:enter-labelx}
\end{figure}

\textit{Generalized Carnot bound--}
We now present a generalized Carnot bound for our setup, showing that coupling a three-level maser to a relativistically moving cold reservoir can enable efficiency beyond the conventional Carnot limit, while sustaining a finite power output. In the high-temperature limit, we analyze families of efficiency–power curves for fixed parameter sets (Fig.~\ref{fig:enter-labelx}). Each curve exhibits the expected trade-off between efficiency and power, with its endpoints at vanishing power marking the maximum efficiency achievable for that set. From the envelope of these endpoints, we extract a \textit{compact analytic expression} for the efficiency bound:
\begin{equation}
    \eta_{\rm up} = 1-\frac{T_c}{T_h}\frac{u}{\sinh{u}}, \label{upperbound}
\end{equation} 
where $u$ denotes the rapidity of the cold bath, and the factor $u/\sinh{u}$  captures the influence of relativistic motion on the upper bound of efficiency. This generalized bound reduces to the Carnot efficiency $\eta_C=1-T_c/T_h$ in the non-relativistic regime $u\rightarrow 0$, thereby seamlessly bridging equilibrium and nonequilibrium relativistic regimes within a unified framework.

The enhancement originates from the relativistic Doppler reshaping of the reservoir spectrum under the Lorentz transformation. In the system’s rest frame, the moving bath no longer appears thermal: photon frequencies are Doppler shifted and their populations acquire an angular dependence \cite{Costa1995}. This anisotropy drives the working medium into a nonthermal steady state, invalidating the assumptions underlying the standard Carnot bound. The form of this upper bound can be justified by assigning an effective temperature,  $T_c^{\rm eff}=T_c u/\sinh{u}$, to the moving cold reservoir. With this effective temperature, the upper bound takes the familiar Carnot form: $\eta_{\rm up} = 1-T_c^{\rm eff}/T_h$. This effective temperature follows from directional averaging \cite{Landsberg1996}. In general, an observer in the system’s rest frame perceives a direction-dependent effective temperature for a moving thermal reservoir, $T^{\theta}_{\rm eff}(T,\theta,u) = T\operatorname{sech}(u)/(1-v\cos\theta)$, with $\theta$ the angle between the motion axis and the line of sight. Averaging over the full solid angle removes the angular dependence, yielding $\langle 1/(1-v\cos\theta)\rangle = u/\tanh{u}$, and hence $T_{\rm eff}=T u/\sinh{u}$.




\begin{figure}
    \centering
    \includegraphics[width=1\linewidth]{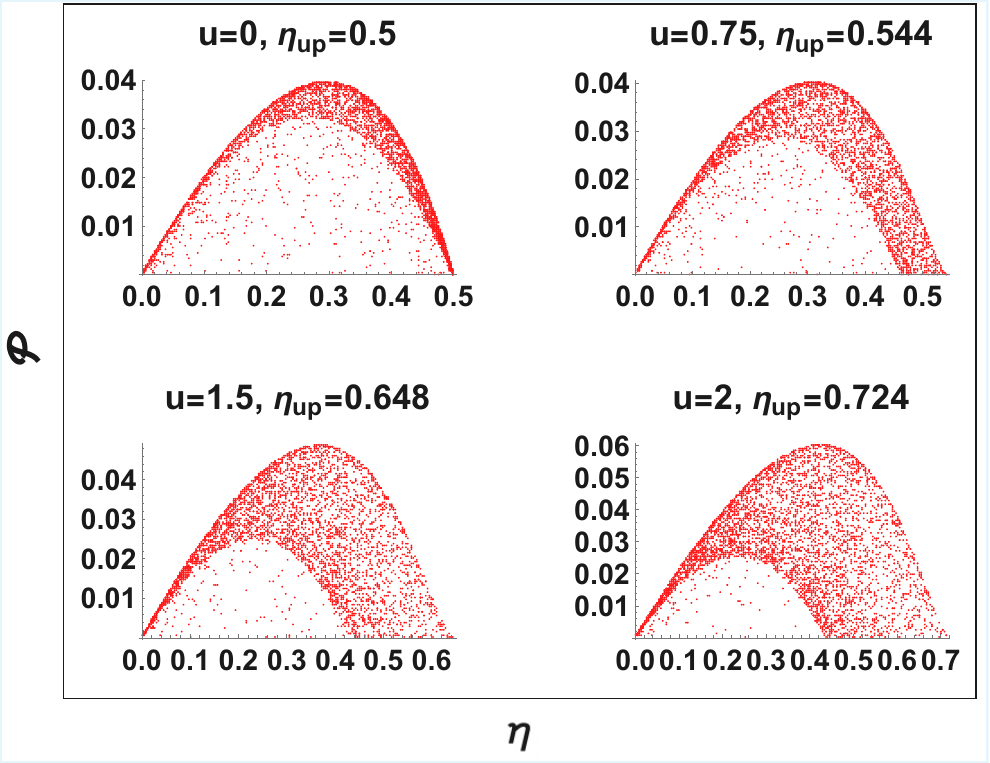}
    \caption{\justifying
Efficiency–power $(\eta,\mathcal{P})$ scatter plots generated via Monte Carlo sampling over random system configurations at fixed temperature ratio $\tau=\beta_h/\beta_c=0.4/0.8$ and relativistic parameter 
$u$  (each panel shows a different $u$. The data illustrate the characteristic efficiency–power trade-off and remain strictly below the   bound
$\eta_{\rm up} = 1 - \tau \, u / \sinh u$, approaching it only as $\mathcal{P}\rightarrow0$. We sample the frequencies uniformly, with $\Omega_c\in [0.01,5]$ and $\Omega_c\in [0.01,10]$; all other parameters are set to unity ($\Gamma_c=\Gamma_h=\Xi=1$). }
    \label{fig:enter-label1}
\end{figure}

Although this efficiency bound is initially extracted from numerical simulations in the high-temperature regime, we find that it remains robust across all temperature regimes. To test its generality, we perform extensive Monte Carlo sampling of efficiency–power pairs across random system configurations at fixed temperature ratio $\tau$ and relativistic parameter $u$ (see Fig. 3). The ($\eta, P$) scatter clearly manifests the efficiency–power trade-off, with all points confined below the bound in Eq.~(\ref{upperbound}) and approaching it only as $\mathcal{P}\rightarrow0$. 
This consistency across different relativistic parameters and temperature regimes shows that the bound is not a high-temperature artifact, but instead reflects a genuine and universal thermodynamic limitation imposed by the relativistic reshaping of the reservoir spectrum.



\begin{figure}[H]
    \centering
    \includegraphics[width=1\linewidth]{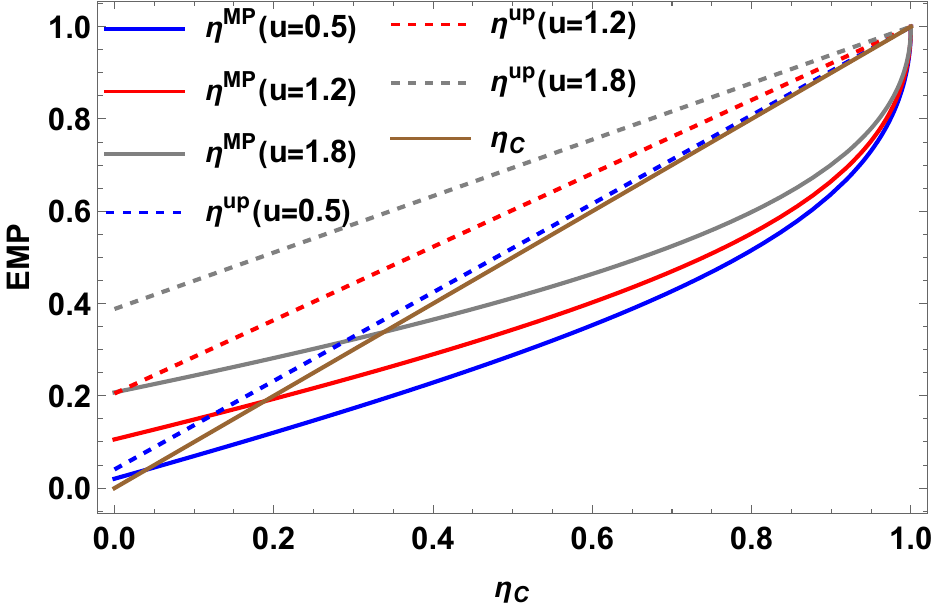}
    \caption{\justifying
Efficiency at maximum power (EMP) plotted as a function of Carnot efficiency $\eta_C = 1 - \tau$ for several fixed values of the relativistic rapidity parameter $u$.  EMP exceeds the conventional Carnot limit over a wide range of $\eta_C$, highlighting the performance advantage enabled by relativistic motion. However, in all cases, the EMP remains strictly bounded above by the generalized bound  $\eta_{\rm up}$ [Eq. (\ref{upperbound})], corroborating the validity of our bound.}

    \label{fig:enter-label2}
\end{figure}

\textit{Efficiency at maximum power--} Having established the limiting efficiency, we now turn to performance at finite power by deriving and analyzing the \textit{efficiency at maximum power} (EMP) in the strong driving ($ \Xi\gg\Gamma_{h,c}$) and  
 in conjunction with the high-temperature limit ($n_{h,c}\gg1$). In this asymptotic regime, the power output reads as:
\begin{equation}
    \mathcal{P} = \frac{\Gamma  \left(\Omega_c - \Omega_h \right) \left(\tau \, u \,\Omega_h - \Omega_c \sinh (u)\right)}{3 \left(\Omega_c \sinh (u)+\tau \, u \,\Omega_h \right)}.
\end{equation}
Here, we used the notation $\Gamma_h=\Gamma_c=\Gamma$. Optimizing the above equation with respect to $\Omega_c$ (keeping $\Omega_h$ fixed), we obtain the following expression for the EMP:
\begin{subequations}
\begin{align}
    \eta^{\rm u}_{\rm MP} 
    = 1 - \left( 
    \sqrt{2 \tau^2 u^2 \, \text{csch}^2(u) + \tau u \, \text{csch}(u)} 
    - \tau u \, \text{csch}(u) 
    \right).
    \label{EMP}
\end{align}
For $u\rightarrow0$, the EMP reduces to
\begin{equation}
\eta_{\rm MP}=2-\eta_C-\sqrt{2(2-3\eta_C+\eta_C^2)}\approx \frac{\eta_C}{2}+\frac{\eta_C^2}{16}+{64}+O(\eta_C^3).
\end{equation}
This coincides with the \textit{Curzon–Ahlborn} result at linear order, 
\begin{eqnarray}
    \eta_{\rm CA}=1-\sqrt{1-\eta_C}\approx \frac{\eta_C}{2}+\frac{\eta_C^2}{16}+O(\eta_C^3),
\end{eqnarray}
\end{subequations}
 but deviates at $O(\eta_C^2)$.
In Fig. \ref{fig:enter-label2}, we plot the EMP as a function of the Carnot efficiency
$\eta_C=1-\tau$	 for several fixed values of the relativistic parameter 
$u$. For each case, we observe that the EMP can exceed the Carnot limit over a broad range of 
$\eta_C$ values. 
Nevertheless, in all instances, the EMP remains strictly bounded from above by the generalized upper bound
$\eta^{\rm up}$ introduced in Eq. (\ref{upperbound}), thereby further reinforcing the \textit{validity} of the proposed bound.

\begin{figure}
    \centering
    \includegraphics[width=1\linewidth]{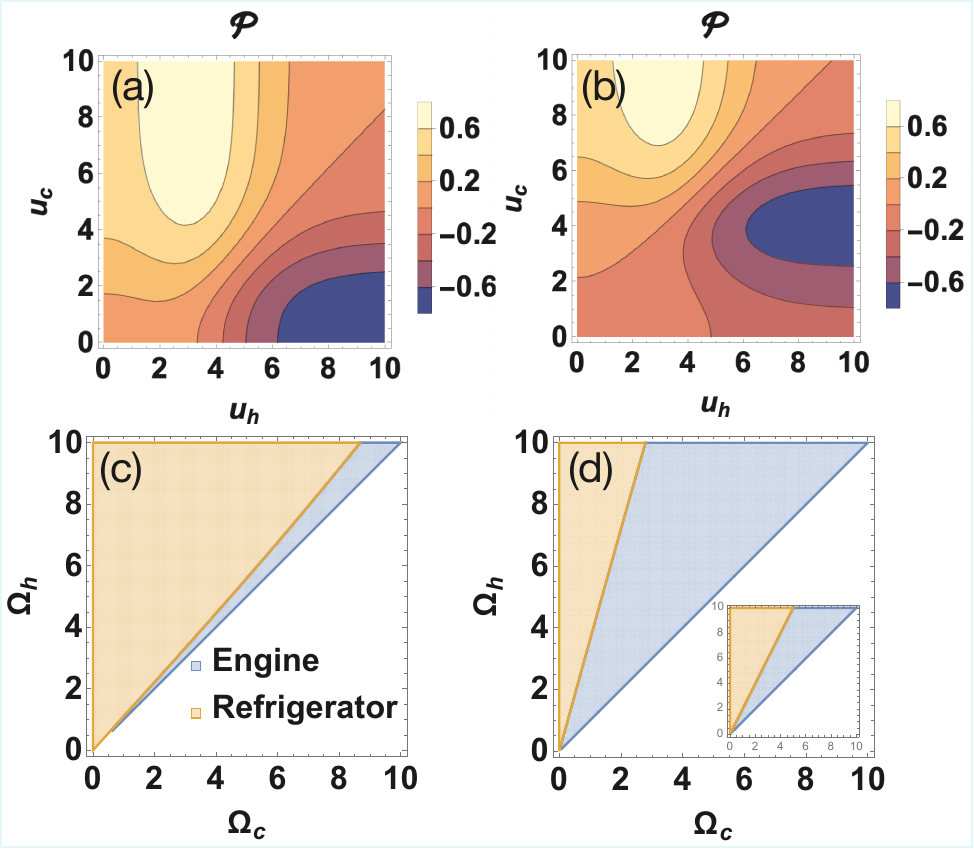}
   \caption{\justifying{Contour plots of power output and operational modes of the thermal machine in different parameter regimes. Panels (a) and (b) show the power output as a function of the reservoir rapidities $u_h$ and $u_c$ for two temperature settings: (a) $\beta_h = 0.01$, $\beta_c = 0.08$; (b) $\beta_h = \beta_c = 0.01$. Panels (c) and (d) display the operation mode, heat engine or refrigerator, in the $(\Omega_c, \Omega_h)$ parameter space for (c) $u_h = 2$, $u_c = 0$, and (d) $u_h = 0$, $u_c = 2$, with fixed inverse temperatures $\beta_h = 0.04$ and $\beta_c = 0.08$. In all panels, the coupling constants are set to $\Gamma_h = \Gamma_c = \Xi = 1$. For panels (a) and (b), the energy parameters are chosen as $\Omega_h = 10$ and $\Omega_c = 5$.}}
    \label{contours}
\end{figure}

\textit{Operation mode transition--} After uncovering how relativistic motion reshapes the fundamental efficiency bounds, we now delve into its profound influence on work extraction and the operational regimes of the thermal machine. Fig. \ref{contours} presents contour plots that encapsulate the thermodynamic behavior of a relativistic QTM in two distinct parameter spaces. The top row, comprising panels (a) and (b), displays the power output as a function of the reservoir rapidities $(u_h, u_c)$, while the bottom row, panels (c) and (d), illustrates the machine's operation mode, either as a heat engine or a refrigerator, in the parameter space  $(\Omega_c, \Omega_h)$ spanned by frequencies of the system.

In panel (a), the parameters are chosen such that, under static conditions, the system would operate solely as a heat engine. Introducing finite reservoir rapidities, however, changes the picture dramatically: the power landscape in the $(u_h, u_c)$ plane becomes strongly asymmetric, with $u_h$ and $u_c$ denoting the rapidities of the hot and cold reservoirs, respectively. As seen in Fig.~\ref{contours}(a), keeping the cold reservoir rapidity fixed and increasing the hot reservoir rapidity from zero to a finite value switches the operation from engine mode to refrigerator mode. Conversely, increasing the cold reservoir rapidity enhances the magnitude of the positive power output. This reveals a non-monotonic dependence of work extraction on reservoir motion, with pronounced differences between moving-hot and moving-cold configurations: a relativistically moving cold reservoir favors heat engine operation, whereas a moving hot reservoir favors refrigeration, showing that relativistic motion can be harnessed to shift the operational boundaries between the two regimes.

A striking consequence of this asymmetry is that work extraction remains possible even when the hot and cold reservoirs are initially at the same temperature. As shown in Fig.~\ref{contours}(b), increasing the cold rapidity gradually triggers a finite power output, and the thermal machine can operate as a heat engine without having any temperature gradient. In this case, it is the motion itself, not a temperature difference, that drives the system out of equilibrium, enabling energy extraction purely from relativistic effects. At first sight, this may seem to contradict the second law of thermodynamics, but as discussed earlier, the effect is fully consistent once the motion-induced spectral modifications are considered, which render the moving reservoir effectively non-thermal.

The thermodynamic operation maps in panels (c) and (d), plotted in the ($\Omega_c,\Omega_h$) space, further illustrate how relativistic motion reshapes the boundaries between engine and refrigerator regimes. When only one reservoir moves, the contrast between moving-hot and moving-cold configurations mirrors the asymmetry seen in panels (a) and (b). The inset of panel (d) shows the case $u_h=u_c$: with both reservoirs moving at the same rapidity, the engine and refrigerator regions remain symmetrically distributed, indicating that the positive-work condition depends only on relative motion. However, the absence of any contour line corresponding to $u_h=u_c$  in panels (a) and (b) shows that absolute motion still influences the magnitude of power extraction, revealing a subtle interplay between relative and absolute motion in determining performance.

\justifying
\color{black}
\textit{Conclusion--} 
Our study demonstrates that relativistic motion fundamentally reshapes the thermodynamic behavior of quantum thermal machines. Through a detailed numerical analysis of the engine's performance, we identified a generalized Carnot bound that reduces to the standard Carnot limit in the stationary-bath case. Remarkably, the efficiency at maximum power can exceed the conventional Carnot efficiency, while always remaining constrained by this generalized bound. In addition, we uncovered pronounced asymmetries between moving-hot and moving-cold configurations, showing that relativistic motion enables positive work extraction even when both reservoirs are at the same temperature. These findings establish relativistic motion as a genuine thermodynamic resource for quantum heat engines.
 
    
    
    
    

\textit{Outlook--} 
Up to now, studies of relativistic heat engines have focused primarily on average thermodynamic quantities, while the impact of relativistic motion on the precision (i.e., fluctuations) of quantum thermal machines remains an open question. A natural next step is to formulate a thermodynamic uncertainty relation (TUR) for relativistically moving QTMs, which would capture the trade-offs among power, fluctuations, and entropy production, thereby clarifying the fundamental cost of precision in motion-assisted processes. In parallel, advancing towards experimental realizations of such heat engines will be crucial for establishing a systematic theory of relativistic quantum thermodynamics and for paving the way toward motion-enabled thermal technologies.

\vspace{3mm}
\textit{Note:} Upon completion of this work, a related study on relativistic quantum heat engines surpassing Carnot efficiency appeared on arXiv \cite{moustos2025surpassingcarnotefficiencyrelativistic}. Their model is based on a two-qubit swap engine, whereas our approach employs an autonomous system as the working medium.  
\vspace{3mm}

\textit{Acknowledgments:} T. P. thanks  Brij Mohan and Manabendra Nath Bera for the fruitful discussions and acknowledges research funding from the QVLS-Q1 consortium, supported by the Volkswagen Foundation and the Ministry for Science and Culture of Lower Saxony. P. C. acknowledges support from the International Postdoctoral Fellowship of the Ben May Center for Theory and Computation. K. C. acknowledges support from the Danish National Research Foundation (bigQ, DNRF). V. S. acknowledges financial support through the KIAS Individual Grant No. PG096801 at the Korea Institute for Advanced Study.


%

\onecolumngrid
\newpage

\appendix

\section{Superoperator and rotating frame representation\label{App.A}}
The dissipative Lindblad superoperator, which encapsulates the non-unitary dynamics arising from the interaction between the system and its thermal reservoirs (hot and cold reservoirs), is formally expressed as:
\begin{eqnarray}\label{eqnA1}
    \mathcal{L}_h [\rho] = \Gamma_h (n_h+1) \left(2 |\mathcal{G}\rangle \bra{\mathcal{G}} \rho_{rr} - \ket{r}\bra{r} \rho - \rho \ket{r}\bra{r} \right) + \Gamma_h n_h (2\ket{r}\bra{r} \rho_{\mathcal{G}\mathcal{G}}-\ket{\mathcal{G}}\bra{\mathcal{G}}  \rho- \rho \ket{\mathcal{G}}\bra{\mathcal{G}}), \\
    \mathcal{L}_c [\rho] = \Gamma_c (n_c+1) \left(2 |\mathcal{G}\rangle \bra{\mathcal{G}} \rho_{ee} - \ket{e}\bra{e} \rho - \rho \ket{e}\bra{e} \right) + \Gamma_c n_c (2\ket{e}\bra{e} \rho_{\mathcal{G}\mathcal{G}}-\ket{\mathcal{G}}\bra{\mathcal{G}}  \rho- \rho \ket{\mathcal{G}}\bra{\mathcal{G}}).
    \end{eqnarray}
 Here $\Gamma_h$ and $\Gamma_c$ are the Weisskoff-Wigner decay constants, $n_{h(c)} = \frac{1}{e^{\Omega/k_B T_{h(c)}}-1}$ is the average population of the photons in the cold(hot) reservoir. 

This superoperator governs the irreversible evolution of the system's density matrix due to its coupling with external environments and plays a central role in capturing the effects of dissipation. It encodes the microscopic processes through which energy is exchanged with the reservoirs and is essential for describing the open-system dynamics that underlie thermodynamic operations such as work extraction and refrigeration. The structure of this operator depends on the spectral properties of the reservoirs, the nature of system-reservoir coupling, and the transition rates determined by detailed balance or generalized fluctuation-dissipation relations.

For this model, it is possible to identify a rotating reference frame, in which the steady-state density matrix, denoted as $\rho_\mathcal{R}$, becomes time-independent. This transformation effectively removes the explicit time dependence associated with the system's interaction with the non-stationary environments, allowing the dynamics to be analyzed in a simplified, co-moving frame. In this rotating frame, the Hamiltonian is
\begin{eqnarray}
    \Tilde{\mathcal{H}}= (\Omega_\mathcal{G} \ket{\mathcal{G}}\bra{\mathcal{G}} + \Omega/2 \ket{r}\bra{r}- \Omega/2 \ket{e}\bra{e}).
\end{eqnarray}
In this rotating frame, the Lindblad superoperators that describe the dissipative interactions with the thermal reservoirs remain invariant under the transformation. The evolution of the system in this rotated frame is depicted as
\begin{eqnarray}\label{A4}
    \dot{\rho}_{\mathcal{R}} = -i[\mathcal{H}_0- \Tilde{\mathcal{H}}+\mathcal{V}_\mathcal{R},\rho_\mathcal{R}] +\mathcal{L}_h^{\rm rel} [\rho_\mathcal{R}] +\mathcal{L}_c^{\rm rel} [\rho_\mathcal{R}],
\end{eqnarray}
where $\mathcal{V}_\mathcal{R} = \Xi (\ket{r}\bra{e}+\ket{e}\bra{r})$, where, $\mathcal{V}_\mathcal{R}$ is the $\mathcal{V}$ the rotating frame where it becomes time-independent.

Incorporating the effects of relativistic motion on the coupling between the system and the reservoirs:

\begin{equation}
    \mathcal{L}_{h}^{\rm rel}[\rho] = \Gamma_h(n_h^{\rm rel}+1)\left[\sigma_{\mathcal{G}r}\rho \sigma_{\mathcal{G}r}^\dagger -\frac{1}{2}\left(\sigma_{\mathcal{G}r}^\dagger \sigma_{\mathcal{G}r}\rho+\rho \sigma_{\mathcal{G}r}^\dagger \sigma_{\mathcal{G}r} \right) \right]  
    +
    \Gamma_h n_h^{\rm rel} \left[\sigma_{\mathcal{G}r}^\dagger\rho \sigma_{\mathcal{G}r} -\frac{1}{2}\left(\sigma_{\mathcal{G}r}\sigma_{\mathcal{G}r}^\dagger \rho+\rho \sigma_{\mathcal{G}r}\sigma_{\mathcal{G}r}^\dagger \right) \right] , \label{disshot}
\end{equation}
\begin{equation}
    \mathcal{L}_{c}^{\rm rel}[\rho] = \Gamma_c(n_c^{\rm rel}+1)\left[\sigma_{\mathcal{G}e}\rho \sigma_{\mathcal{G}e}^\dagger -\frac{1}{2}\left(\sigma_{\mathcal{G}e}^\dagger \sigma_{\mathcal{G}e}\rho+\rho \sigma_{\mathcal{G}e}^\dagger \sigma_{\mathcal{G}e} \right) \right]  
    +
    \Gamma_c n_c^{\rm rel}\left[\sigma_{\mathcal{G}e}^\dagger\rho \sigma_{\mathcal{G}e} -\frac{1}{2}\left(\sigma_{\mathcal{G}e}\sigma_{\mathcal{G}e}^\dagger \rho+\rho \sigma_{\mathcal{G}e}\sigma_{\mathcal{G}e}^\dagger \right) \right]. \label{disscold}
\end{equation}

Note that $\mathcal{L}_h^{\rm rel}[\rho]$ and $\mathcal{L}_c^{\rm rel}[\rho]$ retain the same structure as in Eqs.~(\ref{disshot}) and (\ref{disscold}), with the only difference being that the coefficients representing the mean number of reservoir quanta, $n_h^{\rm rel}$ and $n_c^{\rm rel}$, deviate from the usual Planckian form due to relativistic effects. $n_h^{\rm rel}$ and $n_c^{\rm rel}$ take different forms depending on whether the system-reservoir interaction follows the UDW  coupling.
\color{black}
For a thermal bath in relativistic motion with rapidity $u$, the UDW detector response function yields the mean field occupation number~\cite{PapadatosA,PapadatosB,myers2021quantum}:
\begin{equation}
N_{u_{h(c)}}(\Omega_{h(c)}) = \frac{1}{2\beta \Omega_{h(c)} \sinh u_{h(c)}}\,
\ln\!\left[\frac{1-e^{-\beta \Omega_{h(c)} e^{|u_{h(c)}|}}}{1-e^{-\beta \Omega_{h(c)} e^{-|u_{h(c)}|}}}\right].
\end{equation}

which reduces to the Planck distribution at $u=0$. This quantity enters the dissipators of Eq.~\eqref{eq:master}. In the rotating frame, the hot- and cold-bath superoperators take the form
\begin{align}
\mathcal{L}_h^{\mathrm{rel}}[\rho] &= 
\Gamma_h\big(N_h(u_h)+1\big)\,\mathcal{D}[\sigma_{\mathcal{G}r}]\rho
+ \Gamma_h N_h(u_h)\,\mathcal{D}[\sigma_{r\mathcal{G}}]\rho, \\
\mathcal{L}_c^{\mathrm{rel}}[\rho] &= 
\Gamma_c\big(N_c(u_c)+1\big)\,\mathcal{D}[\sigma_{\mathcal{G}e}]\rho
+ \Gamma_c N_c(u_c)\,\mathcal{D}[\sigma_{e\mathcal{G}}]\rho,
\end{align}

where $\mathcal{D}[A]\rho = A\rho A^\dagger - \tfrac{1}{2}\{A^\dagger A,\rho\}$. 
\color{black}

\section{Steady-State Solution of Density Matrix Equations for a Three-Level Heat Engine \label{App. B}}

The steady-state density matrix of the three-level quantum heat engine is presented below. By systematically substituting the explicit forms of the bare Hamiltonian $\mathcal{H}_0$, $\bar{\mathcal{H}}$, and the residual interaction term $\mathcal{V}_R$ into Eq.~\eqref{eq:master}, we derive the corresponding set of time-evolution equations governing the dynamics of the density matrix elements. The time evolution equations for the density matrix elements are derived as follows:
\begin{equation}
\dot{\rho_{11}} = i\Xi(\rho_{10} - \rho_{01}) - 2\Gamma_h \left[(n_h+1)\rho_{11} - n_h\rho_{gg}\right],
\end{equation}
\begin{equation}
\dot{\rho_{00}} = -i\Xi(\rho_{10} - \rho_{01}) - 2\Gamma_c \left[(n_c+1)\rho_{00} - n_c\rho_{gg}\right],
\end{equation}
\begin{equation}
\dot{\rho_{10}} = -\left[\Gamma_h(n_h+1) + \Gamma_c(n_c+1)\right]\rho_{10} + i\Xi(\rho_{11} - \rho_{00}),
\end{equation}
\begin{equation}
\rho_{11} = 1 - \rho_{00} - \rho_{gg},
\end{equation}
\begin{equation}
\dot{\rho_{01}} = \dot{\rho_{10}^\ast}.
\end{equation}

To find the steady-state solution, we set \(\dot{\rho_{mn}} = 0\) (for \(m, n = 0, 1\)) in Eq.~\eqref{eqnA1} to \eqref{A4}. This yields the following solution for \(\rho_{10}\):

\begin{equation}
\rho_{10} = \frac{2i\Xi(n_h-n_c)\Gamma_c\Gamma_h}
{
4\Xi^2[(1+3n_h)\Gamma_h + (1+3n_c)\Gamma_c] + \Gamma_c\Gamma_h[1+2n_h+n_c(2+3n_h)][(1+n_c)\Gamma_c + (1+n_h)\Gamma_h ] 
}.
\end{equation}

The corresponding element \(\rho_{01}\) is given by:

\begin{equation}
\rho_{01} = \rho_{10}^\ast.
\end{equation}

Using the trace of the density matrix, the output power \(P\) and heat flow \(\dot{Q_h}\) are expressed as:

\begin{equation}
P = i\hbar\Xi(\Omega_h-\Omega_c)(\rho_{01} - \rho_{10}),
\end{equation}
\begin{equation}
\dot{Q_h} = i\hbar\Xi\Omega_h(\rho_{10} - \rho_{01}).
\end{equation}
Substituting Eqs. (B6) and (B7) in Eqs. (B8) and (B9), the final expressions for power and efficiency gven by Eqs. (\ref{power}) and (\ref{eff}) in the main text, can be obtained.
\color{black}  

\color{black} 
\color{black}
\section{Monte Carlo Estimation of the Generalized Bound}\label{App.MC}

To extract the generalized Carnot-like efficiency bound in relativistic QTMs, we perform a Monte Carlo sampling over a large set of parameters defining the system and reservoir configurations. For fixed temperature ratio $\tau = \beta_h/\beta_c$ and rapidity parameter $u$, we randomly sample $10^5$ combinations of transition frequencies $(\Omega_c, \Omega_h) \in [1, 10]$, inverse temperatures $(\beta_c, \beta_h)$, and compute the power output $\mathcal{P}$ and efficiency $\eta$. 

The resulting $(\eta, \mathcal{P})$ pairs are plotted and analyzed using convex hull methods to identify the outer performance boundary. The point where power vanishes gives the maximum efficiency under reversible conditions. Repeating this over various $u$ reveals a universal upper bound:
\[
\eta^{\rm up}(u) = 1 - \tau \frac{u}{\sinh u},
\]
which we interpret as a generalized Carnot limit in relativistic settings. This bound holds robustly across both high- and moderate-temperature regimes and captures the limiting efficiency beyond stationarity.

\vspace{0.5em}
\noindent\textbf{Pseudocode: Monte Carlo Sampling for $\eta^{\rm up}$}
\begin{flushleft}
\begin{tabbing}
\hspace{0.5cm} \= \textbf{for} $i = 1$ to $N_{\rm samples}$ \textbf{do} \\
\> \quad Sample $\Omega_c$, $\Omega_h$, $\beta_c$, $\beta_h$ at fixed $\tau$, $u$ \\
\> \quad Compute $N_{c,h}(u)$ and $\rho_{\rm ss}$ \\
\> \quad Compute $\mathcal{P}_i$ and $\eta_i$ \\
\textbf{end} \\
Apply convex hull to $(\eta_i, \mathcal{P}_i)$ \\
Extract $\eta^{\rm up}(u)$ at $\mathcal{P} \to 0$
\end{tabbing}
\end{flushleft}


\end{document}